\documentstyle[12pt]{article}
\begin{document}

\def\ds{\displaystyle}
\def\beq{\begin{equation}}
\def\eeq{\end{equation}}
\def\bea{\begin{eqnarray}}
\def\eea{\end{eqnarray}}
\def\beeq{\begin{eqnarray}}
\def\eeeq{\end{eqnarray}}
\def\ve{\vert}
\def\vel{\left|}
\def\ver{\right|}
\def\nnb{\nonumber}
\def\ga{\left(}
\def\dr{\right)}
\def\aga{\left\{}
\def\adr{\right\}}
\def\lla{\left<}
\def\rra{\right>}
\def\rar{\rightarrow}
\def\nnb{\nonumber}
\def\la{\langle}
\def\ra{\rangle}
\def\ba{\begin{array}}
\def\ea{\end{array}}
\def\tr{\mbox{Tr}}
\def\ssp{{\Sigma^{*+}}}
\def\sso{{\Sigma^{*0}}}
\def\ssm{{\Sigma^{*-}}}
\def\xis0{{\Xi^{*0}}}
\def\xism{{\Xi^{*-}}}
\def\qs{\la \bar s s \ra}
\def\qu{\la \bar u u \ra}
\def\qd{\la \bar d d \ra}
\def\qq{\la \bar q q \ra}
\def\gGgG{\la g^2 G^2 \ra}
\def\q{\gamma_5 \not\!q}
\def\x{\gamma_5 \not\!x}
\def\g5{\gamma_5}
\def\sb{S_Q^{cf}}
\def\sd{S_d^{be}}
\def\su{S_u^{ad}}
\def\ss{S_s^{??}}
\def\sbp{{S}_Q^{'cf}}
\def\sdp{{S}_d^{'be}}
\def\sup{{S}_u^{'ad}}
\def\ssp{{S}_s^{'??}}
\def\sig{\sigma_{\mu \nu} \gamma_5 p^\mu q^\nu}
\def\fo{f_0(\frac{s_0}{M^2})}
\def\ffi{f_1(\frac{s_0}{M^2})}
\def\fii{f_2(\frac{s_0}{M^2})}
\def\O{{\cal O}}
\def\sl{{\Sigma^0 \Lambda}}
\def\es{\!\!\! &=& \!\!\!}
\def\ap{\!\!\! &\approx& \!\!\!}
\def\ar{&+& \!\!\!}
\def\ek{&-& \!\!\!}
\def\kek{\!\!\!&-& \!\!\!}
\def\cp{&\times& \!\!\!}
\def\se{\!\!\! &\simeq& \!\!\!}
\def\eqv{&\equiv& \!\!\!}
\def\kpm{&\pm& \!\!\!}
\def\kmp{&\mp& \!\!\!}


\def\simlt{\stackrel{<}{{}_\sim}}
\def\simgt{\stackrel{>}{{}_\sim}}


\title{
         {\Large
                 {\bf
Electromagnetic form factors of the $\rho$ meson in
light cone QCD sum rules
                 }
         }
      }

\author{\vspace{1cm}\\
{\small T. M. Aliev \thanks
{e-mail: taliev@metu.edu.tr}\,\,,
M. Savc{\i} \thanks
{e-mail: savci@metu.edu.tr}} \\
{\small Physics Department, Middle East Technical University,
06531 Ankara, Turkey} } 
\date{}

\begin{titlepage}
\maketitle
\thispagestyle{empty}

\begin{abstract}
We investigate the electromagnetic form factors of the $\rho$ meson in light
cone QCD sum rules. We find that the ratio of the magnetic and charge form
factors is larger than two at all values of $Q^2,~(Q^2 \ge 0.5~GeV^2)$. The
values of the individual form factors at fixed values of $Q^2$ predicted by
the light cone QCD sum rules are quite different compared to the results of
other approaches. These results can be checked in future, when more precise 
data on $\rho$ meson form factors is available.
\end{abstract}

~~~PACS numbers: 12.38.Lg, 13.40.Gp
\end{titlepage}

\section{Introduction}
The QCD sum rules method \cite{R6401} is one of the most powerful tools in
studying low energy hadron physics. In this method, physically measurable
quantities of hadrons are connected with QCD parameters, where hadrons are
represented by their interpolating quark currents. The main idea of this
method is to calculate the correlator functions of the interpolating quark
currents in the deep Eucledian region with the help of operator product
expansion (OPE) which allows one to take into account both perturbative and
nonperturbative contributions. Relevant physical quantities are determined
by matching the correlator to its phenomenological representation.

In the current literature, a new, widely discussed alternative to
traditional sum rules, namely, QCD light cone sum rules (QLCSR) is a
convenient tool for the study of exclusive processes. This method is based
on OPE on the light cone, which is an expansion over twists of the
operators, rather than dimensions, as is the case in the traditional QCD sum
rules. Moreover, in this method, all the nonperturbative dynamics encoded 
in the light cone distribution amplitudes, determine the matrix
elements of the nonlocal operators between the vacuum and and the hadronic
states (more about this method and its applications can be found in
\cite{R6402,R6403}).

In the present work, we study the $\rho$ meson form factors in QLCSR. It
should be mentioned here that the $\rho$ meson form factors are calculated
at intermediate momentum transfer by using the three--point QCD sum rules
method in \cite{R6404}. Recently, $\rho$ meson form factors, including
next--to--leading--order perturbation theory, are analyzed within the same
framework \cite{R6405}. The $Q^2=0$ point lies outside the applicability 
region of the three--point QCD sum rules. Consequently, extrapolating these 
form factors to $Q^2=0$, in principle, one can determine static 
characteristics of the $\rho$ meson, such as magnetic and quadrapole moments.
The $\rho$ meson magnetic moment has already been investigated in 
the framework of the traditional three--point and QLCSR in \cite{R6406} and 
\cite{R6407}, respectively. It should be noted here that, the QLCSR is 
successfully applied to a wide range of problems of the hadron physics, for 
example, magnetic moments of the octet and decuplet baryons are calculated 
in \cite{R6408} and \cite{R6409}, and magnetic moment of the nucleon in 
\cite{R6410}, respectively. 

The paper is organized as follows. In section 2, we derive the sum rules for
the $\rho$ meson electromagnetic form factors. In section 3, our numerical
results and a comparison of them with the results of various approaches 
existing in the literature is presented. 

\section{Theoretical framework} 

In order to determine $\rho$ meson electromagnetic form factors we will use 
QLCSR method. For this purpose, we consider the following correlator function
\bea
\label{e6401}
\Pi_{\mu\nu}(p,q) = i \int d^4x e^{iqx} \la \rho(p) \ve
J_\mu^{el} (x) J_\nu (0) \ve 0 \ra~,
\eea
where 
\bea
J_\nu(x) \es \bar{u}(x) \gamma_\nu d(x)~, \nnb \\
J_\mu^{el}(x) \es \sum_{q=u,d} e_q \bar{q}(x) \gamma_\mu q(x)~, \nnb
\eea
are the $\rho$ meson interpolating and electromagnetic currents,
respectively. 

Firstly, we calculate the phenomenological part of the correlator. By
inserting the complete set of states which have the  quantum numbers of the
$\rho$ meson, between the currents in Eq. (\ref{e6401}), we get
\bea
\label{e6402}
\Pi_{\mu\nu} = \frac{\ds \la \rho(p,\varepsilon) \ve
J_\mu^{el}(x) \ve \rho(p^\prime,\varepsilon^\prime) \ra \la
\rho(p^\prime,\varepsilon^\prime)\ve J_\nu (0) \ve 0 \ra}
{\ds p^{\prime 2} - m_\rho^2} + \cdots~,
\eea
where $p^\prime = p+q$, $q$ is the momentum of the electromagnetic
current, $\varepsilon$ is the polarization of the $\rho$ meson and $\cdots$
describes contributions from higher states. The matrix element
$\la \rho \ve j_\nu \ve 0 \ra$ is determined as
\bea
\label{e6403}
\la \rho \ve j_\nu \ve 0 \ra  = f_\rho m_\rho \varepsilon_\nu^\ast (p)~.
\eea
 Assuming parity and time--reversal invariance, the electromagnetic vertex
of the $\rho$ meson can be written in terms of three Lorentz invariant
form factors \cite{R6411}
\bea
\label{e6404}
\la \rho(p,\varepsilon) \ve J_\mu^{el} \ve \rho(\rho^\prime,\varepsilon^\prime)
\ra \es \varepsilon^{\ast\alpha} \varepsilon^{\prime\beta}
\Big\{ - G_1(Q^2) g_{\alpha\beta}
(p+p^\prime)_\mu - G_2(Q^2) (q_\beta g_{\mu\alpha} - 
q_\alpha g_{\mu\beta}) \nnb \\
\ar \frac{1}{2 m_\rho^2} G_3(Q^2) q_\alpha q_\beta (p+p^\prime)_\mu \Big\}~,
\eea
where $Q^2=-q^2$ is the square of the momentum transfer. 
It should be mentioned here that, in practical computation, instead of
calculating the Lorentz invariant form factors $G_i(Q^2)$, the physical 
charge $G_C$, the magnetic $G_M$ and quadrapole $G_Q$ form factors 
are often used.

The Lorentz invariant form factors $G_i(Q^2)$ are related to the charge,
magnetic and quadrapole form factors through the following relations
\bea
\label{e6405}
G_C \es G_1 + \frac{2}{3} \eta F_{\cal D}~, \nnb \\
G_M \es G_2~, \nnb \\
G_Q \es G_1 - G_2 + (1+\eta) G_3~,
\eea
where $\eta = Q^2/4 m_\rho^2$. At zero momentum transfer, these form factors
are proportional to the usual static quantities of charge $e$, magnetic 
moment $\mu$ and quadrapole moment ${\cal D}$
\bea
\label{e6406}
eG_C(0) \es e~,\nnb \\
eG_M(0) \es 2 m_\rho \mu~, \nnb \\
eG_Q(0) \es m_\rho^2 {\cal D}~.
\eea
Substituting Eqs. (\ref{e6403}) and (\ref{e6404}) into Eq, (\ref{e6402}),
the phenomenological part of the correlator takes the following form
\bea
\label{e6407}
\Pi_{\mu\nu} \es \frac{f_\rho m_\rho}{{p^\prime}^2-m_\rho^2}
\Bigg\{ G_1 (p+p^\prime)_\mu
\Bigg[\varepsilon_\nu - \frac{(\varepsilon q) p^\prime_\nu}{m_\rho^2}\Bigg] +   
G_2  \Bigg[\varepsilon_\mu q_\nu - (\varepsilon q) g_{\mu\nu} \nnb \\
\ar \frac{1}{m_\rho^2} \Bigg( 
\frac{1}{2} Q^2 \varepsilon_\mu p^\prime_\nu + (q \varepsilon)
p^\prime_\mu p^\prime_\nu\Bigg) \Bigg]
-  G_3 \frac{(p+p^\prime)_\mu}{2 m_\rho^2} (q \varepsilon) \Bigg( q_\nu + 
\frac{Q^2}{2 m_\rho^2} p_\nu^\prime \Bigg) \Bigg\}~.
\eea
It is apparent from Eq. (\ref{e6407}) that there are many different structures
each of which can be used to extract the above--mentioned form factors.

In order to find out the $\rho$ meson electromagnetic form factors, we  
pick the following three structures: $\varepsilon_\nu p_\mu$,
$\varepsilon_\mu q_\nu$ and $(\varepsilon q)p_\mu q_\nu$. Hence we can
write
\bea
\label{e6408}
\Pi_{\mu\nu}(p,q) \es \Pi_1(p,q) \varepsilon_\nu p_\mu+ \Pi_2(p,q)
\varepsilon_\mu q_\nu + \Pi_3(p,q) (\varepsilon q)p_\mu q_\nu +
\cdots~,
\eea
where
\bea
\label{e6409}
\Pi_1(p,q) \es - G_1(Q^2) \frac{2 f_\rho m_\rho}{m_\rho^2 - (p+q)^2}~,\nnb \\
\Pi_2(p,q) \es - G_2(Q^2) ( 1 + 2 \eta) 
\frac{f_\rho m_\rho}{m_\rho^2 - (p+q)^2}~, \nnb\\
\Pi_3(p,q) \es [2 G_1(Q^2) - G_2(Q^2) + G_3(Q^2) (1 +
2\eta)] \frac{f_\rho m_\rho}{m_\rho^2[m_\rho^2 -
(p+q)^2]}~.
\eea 
For the invariant amplitudes $\Pi_i(p,q)$, one can write a general
dispersion relation in $(p+q)^2$ in the form
\bea
\label{e6410}
\Pi_i(p,q) = \int ds \frac{\rho_i(s,q)}{s-(p+q)^2} + \mbox{\rm subtr.}~,
\eea
where the spectral density corresponding to Eq. (\ref{e6407}) is
\bea
\label{e6411}
\rho_i(s,q) = A F_i(Q^2) \delta(s-m_\rho^2)~,
\eea
where
\bea
F_1 \es - 2 G_1(Q^2) f_\rho m_\rho~,\nnb \\
F_2 \es - G_2(Q^2) ( 1 + 2 \eta) f_\rho m_\rho~,\nnb \\
F_3 \es [2 G_1(Q^2) - G_2(Q^2) + G_3(Q^2) (1 +
2 \eta)] \frac{f_\rho m_\rho}{m_\rho^2}~.\nnb
\eea

According to QCD sum rules philosophy, in constructing sum rules for the 
form factors we need representation of the correlator function from QCD
side. After contracting quark fields, the correlator function takes the
form:
\bea
\label{e6412}
\Pi_{\mu\nu} \es \int d^4x e^{iqx} \la \rho(p) \ve e_u \bar{u}(x)
\gamma_\mu S_d(x,0) \gamma_\nu d(0)
+ e_d \bar{u}(0) \gamma_\nu S_u(0,x) \gamma_\mu d(x) \ve 0 \ra~,
\eea
where $S_q(x)$ is the full quark operator of $u$ (or $d$) quark. Imposing
$SU(2)$ symmetry and neglecting $u$ and $d$ quark masses, we have 
$S_u(x,0)=S_d(x,0)=S(x)$, and hence the full light quark propagator takes the
form
\bea
\label{e6413}   
S(x) \es \frac{i \not\!x}{2 \pi^2 x^4} - \frac{\la \bar{q}q\ra}{12} - 
\frac{x^2 m_0^2}{192} \la \bar{q}q\ra
- i g_s \int_0^1 dv \Bigg[ \frac{\not\!x \sigma_{\alpha\beta}}
{16 \pi^2 x^2} G_{\alpha\beta}(vx) - \frac{i}{4 \pi^2 x^2}
v x_\alpha  G_{\alpha\beta} \gamma_\beta\Bigg]~.
\eea

Note that the second and third terms of $S(x)$ do not give any contribution 
to the considered problem after Borel transformation is carried out 
(see below). Rewriting Eq. (\ref{e6413}) in momentum representation and 
substituting it in Eq. (\ref{e6412}), we get
\bea
\label{e6414}
\Pi_{\mu\nu} \es \frac{i^2}{4} \int d^4x d^4k e^{i(q-k)x} \Bigg\{
e_u \la \rho(p) \ve \bar{u}(x) \Gamma_i d(0) \ve 0 \ra \mbox{\rm Tr} \gamma_\mu
\frac{\not\!{k}}{k^2}\gamma_\nu \Gamma_i \nnb \\
\ek e_u \la \rho(p) \ve \bar{u}(x) g G_{\alpha\beta}(vx)\Gamma_i d(0) \ve 0 \ra 
\mbox{\rm Tr} \gamma_\mu \Bigg[\frac{1}{2} \frac{\not\!{k}}{k^4}
\sigma_{\alpha\beta} - \frac{v}{k^2} x_\alpha \gamma_\beta \Bigg] 
\gamma_\nu\Gamma_i \nnb \\
\ar e_d \la \rho(p) \ve \bar{u}(0) \Gamma_i d(x) \ve 0 \ra \mbox{\rm Tr}
\gamma_\nu \frac{\not\!{k}}{k^2}\gamma_\mu \Gamma_i \nnb \\
\ek e_d \la \rho(p) \ve \bar{u}(0) g G_{\alpha\beta}(vx) \Gamma_i d(x) \ve 0 \ra 
\mbox{\rm Tr} \gamma_\nu \Bigg[\frac{1}{2} \frac{\not\!{k}}{k^4}
\sigma_{\alpha\beta} - \frac{v}{k^2} x_\alpha \gamma_\beta \Bigg] 
\gamma_\mu\Gamma_i \Bigg\}~,
\eea 
where $\Gamma_i= (I,\gamma_5,\gamma_\alpha,\gamma_\alpha\gamma_5,
\sigma_{\alpha\beta})$ is the full set of the Dirac matrices. It follows
from this expression that, in order to calculate the correlator from QCD
side one needs to know the matrix element of the nonlocal operators between
vacuum and the $\rho$ meson, i.e., $\rho$ meson distribution amplitudes (or
wave functions). We can see from Eq. (\ref{e6414}) that main contribution to
the correlator comes from only from the wave functions that contain
odd--number of $\gamma$--matrices that are defined in the following way
\cite{R6414}:
\bea
\label{e6415}
\la \rho (p,\varepsilon) \ve \bar{u}(x) \gamma_\mu d(0) \ve 0 \ra \es
f_\rho m_\rho \Bigg\{ \frac{\varepsilon x}{px} \int_0^1 dv e^{ivpx}
\Bigg[ \phi_\parallel (v,\mu) + \frac{m_\rho^2 x^2}{16} A(v,\mu) 
\Bigg] p_\mu \nnb \\
\ar \Bigg( \varepsilon_\mu - \frac{\varepsilon x}{px} p_\mu \Bigg) \int_0^1 dv
e^{ivpx} g_\perp^v (v,x) \nnb \\
\ek \frac{1}{2} x_\mu \frac{\varepsilon x}{(px)^2}
m_\rho^2 \int_0^1 dv e^{ivpx} C(v,\mu) \Bigg\}~, \\ \nnb \\
\label{e6416}
\la \rho (p,\varepsilon) \ve \bar{u}(x) \gamma_\mu \gamma_5 d(0) \ve 0 \ra \es
\frac{1}{4} f_\rho m_\rho \epsilon_{\mu\alpha\beta\delta} \varepsilon_\alpha
p_\beta x_\delta \int_0^1 dv e^{ivpx} g_\perp^a (v,\mu)~, \\ \nnb \\
\label{e6417}
\la \rho (p,\varepsilon) \ve \bar{u}(x) g G_{\mu\nu} \gamma_\alpha  d(0) \ve 0 \ra
\es -i f_\rho m_\rho P_\alpha \Big( P_\nu \varepsilon_\mu^\perp - 
P_\mu \varepsilon_\nu^\perp\Big) {\cal V} \nnb \\
\ek i f_\rho m_\rho^3 \frac{\varepsilon x}{px} \Big( P_\mu g_{\alpha\nu}^\perp -
P_\nu g_{\alpha\mu}^\perp \Big) \Phi \nnb \\
\ek  i f_\rho m_\rho^3 \frac{\varepsilon x}{(px)^2} P_\alpha \Big( P_\mu x_\nu -
P_\nu x_\mu \Big) \Psi~, \\ \nnb \\
\label{e6418}
\la \rho (p,\varepsilon) \ve \bar{u}(x) g \widetilde{G}_{\mu\nu} \gamma_\alpha
\gamma_5 d(0) \ve 0 \ra \es f_\rho m_\rho P_\alpha \Big( P_\nu
\varepsilon_\mu^\perp - P_\mu \varepsilon_\nu^\perp\Big) A \nnb \\
\ar i f_\rho m_\rho^3 \frac{\varepsilon x}{px} \Big( P_\mu
g_{\alpha\nu}^\perp - P_\nu g_{\alpha\mu}^\perp \Big) \widetilde{\Phi} \nnb \\
\ar f_\rho m_\rho^3 \frac{\varepsilon x}{(px)^2} P_\alpha \Big( P_\mu x_\nu
- P_\nu x_\mu \Big) \widetilde{\Psi}~,
\eea  
where $P_\mu$, $\varepsilon_\mu^\perp$ and $g_{\mu\nu}^\perp$ are defined as
\bea
P_\mu \es p_\mu - \frac{m_\rho^2}{2 px} x_\mu~, \nnb \\
\varepsilon_\mu^\perp \es \varepsilon_\mu - \frac{\varepsilon x}{px} 
\Bigg( P_\mu - \frac{m_\rho^2}{px} x_\mu \Bigg)~,\nnb \\
g_{\mu\nu}^\perp \es \Bigg( g_{\mu\nu} - \frac{P_\mu x_\nu + P_\nu x_\nu}{px}
\Bigg)~,\nnb   
\eea
where $px = Px$ has been used.

In Eqs. (\ref{e6415})--(\ref{e6418}), $\phi_\parallel(v,\mu)$ is the leading
twist--2 wave function, while $g_\perp^v$, $g_\perp^a$ and ${\cal V}$ are
the twist--3 and all the remaining ones are the twist--4 wave functions. In
Eqs. (\ref{e6417}) and (\ref{e6418}) the following relation is used
\bea
{\cal V}(v,px) = \int {\cal D}\alpha e^{ipx(\alpha_1 + v \alpha_3)} {\cal V}
(\alpha_1,\alpha_2,\alpha_3)~,\nnb
\eea
where
\bea
{\cal D}\alpha = d\alpha_1 d\alpha_2 d\alpha_3
\delta(1-\alpha_1-\alpha_2-\alpha_3)~.\nnb
\eea

In order to suppress the contributions of higher states and continuum and
further eliminate the subtraction terms in the dispersion relation, it is
necessary to perform Borel transformation with respect to $-(p+q)^2$. The
contributions of higher states and continuum are subtracted using the quark
hadron duality (for details, see \cite{R6402,R6403},
\cite{R6408}--\cite{R6410}, \cite{R6412} and \cite{R6413}). 

The sum rules for form factors can be obtained after applying Borel
transformation to the two different representation of the invariant
functions $F_i$ and then matching these results, and in doing so, we get
\bea
\label{e6419}
\lefteqn{
G_1(Q^2) = \frac{1}{2} e^{m_\rho^2/M^2} \Bigg\{ \int_{v_0}^1 dv
e^{-s(v)/M^2} \Bigg[ -\frac{m_\rho^2}{4} \frac{1}{v^2 M^2} 
\Big( e_u A^i(v) - e_d A^i(\bar{v}) \Big) } \nnb \\
\ar \frac{1}{v} \Big( e_u \Phi^i(v) - e_d \Phi^i(\bar{v}) \Big)
- \frac{2 m_\rho^2}{v M^2} \Big( e_u C^{ii}(v) - 
e_d C^{ii}(\bar{v}) \Big) \nnb \\
\ar \Big( e_u g_\perp^v(v) - e_d g_\perp^v(\bar{v}) \Big)
+ \frac{1}{4} \Bigg( \frac{1}{v} - \frac{Q^2 + m_\rho^2 v^2}{v^2 M^2}
\Bigg) \Big(e_u g_\perp^a(v) + e_d g_\perp^a(\bar{v}) 
\Big) \Bigg] \Bigg\}~, \\ \nnb \\
\label{e6420}
\lefteqn{
G_2(Q^2) = \frac{2 m_\rho^2}{Q^2 + 2 m_\rho^2} e^{m_\rho^2/M^2} \Bigg\{ 
\int_{v_0}^1 dv e^{-s(v)/M^2} \Bigg[ -\frac{2 m_\rho^2}{v^2 M^2} 
\Big( e_u C^{ii}(v) - e_d C^{ii}(\bar{v}) \Big)} \nnb \\
\ar \frac{1}{v} \Big( e_u g_\perp^v(v) - e_d g_\perp^v(\bar{v}) \Big) 
-\frac{1}{4 v^2} \Bigg( 1 - \frac{Q^2 + m_\rho^2 v^2}{v M^2} \Bigg)
\Big(e_u g_\perp^a(v) + e_d g_\perp^a(\bar{v}) \Big) \Bigg] 
\Bigg\}~, \\ \nnb \\
\label{e6421}
\lefteqn{
-2 G_1 + G_2 - G_3 ( 1 + 2\eta ) = m_\rho^2 e^{m_\rho^2/M^2}
\Bigg\{ \int_{v_0}^1 dv e^{-s(v)/M^2} \Bigg[ -\frac{m_\rho^2}{2 v^3 M^4}
\Big( e_u A^i(v)} \nnb \\
\ek e_d A^i(\bar{v}) \Big) - \frac{1}{2 v^2 M^2} \Big(e_u
g_\perp^a(v) + e_d g_\perp^a(\bar{v}) \Big) + \frac{2}{v^2 M^2} 
\Big( e_u \Phi^i(v) - e_d \Phi^i(\bar{v}) \Big) \nnb \\
\ek \frac{4 m_\rho^2 v }{ v^3 M^4} \Big( e_u C^{ii}(v) - e_d
C^{ii}(\bar{v}) \Big)\Bigg] \Bigg\}~,  
\eea
where $M^2$ is the Borel parameter and
\bea
s(v) \es \frac{(Q^2 + m_\rho^2 v) \bar{v}}{v}~,\nnb \\
\bar{v} \es 1-v~, \nnb \\
v_0 \es -\frac{1}{2 m_\rho^2} \Big[(s_0+Q^2-m_\rho^2)^2 -
\sqrt{(s_0+Q^2-m_\rho^2)^2+4 m_\rho^2 Q^2}\Big]~,\nnb
\eea
where $s_0$ is the continuum threshold.

In Eqs. (\ref{e6419})--(\ref{e6421}), functions $A^i$, $\Phi^i$ and $C^{ii}$  
are defined as follows
\bea
A^i \es - \int_0^v du A(u)~,\nnb \\
\Phi^i(v)  \es - \int_0^v du \Phi(u)~,\nnb \\
C^i(v)  \es - \int_0^v du C(u)~,\nnb \\
C^{ii}(v)  \es - \int_0^v du C^i(u)~. \nnb
\eea

Note that the terms involving three--particle $\rho$ meson wave functions 
are not presented in Eqs. (\ref{e6419})--(\ref{e6421}) since their
expressions are rather lengthy, however, their contributions are taken into
account in the numerical analysis, which constitute about $5\%$ of the total
result.

\section{Numerical analysis}

In this section we present our numerical calculations on charge, magnetic
and quadrapole form factors. The main input parameters of the QLCSR in
regard to the above--mentioned form factors are the $\rho$ meson wave
functions, whose explicit expressions are given in \cite{R6412} and
\cite{R6414}, and we use them in our analysis.

Apart from the wave functions, the sum rules for the form factors depend on
the value of the continuum threshold $s_0$ and Borel parameter $M^2$. In the
present work, we calculate the form factors at three different values of the
threshold, i.e., $s_0=1.8~GeV^2$, $s_0=2.0~GeV^2$ and $s_0=2.2~GeV^2$ (see
also \cite{R6404}).

In Figs. (1), (2) and (3) we present the dependence of $G_C$, $G_M$ and
$G_{\cal D}$ on $M^2$ at six different values $Q^2=0.5~GeV^2$, $Q^2=1.0~GeV^2$,
$Q^2=2.0~GeV^2$, $Q^2=3.0~GeV^2$, $Q^2=4.0~GeV^2$ and $Q^2=5.0~GeV^2$ 
of the momentum transfer, respectively. The Borel parameter $M^2$ in
the sum rule is an auxiliary parameter and therefore physical quantities
must be independent of it. For this reason we must determine the region of
$M^2$ in which the form factors are independent of its value. In determining
the working region of of the Borel parameter $M^2$, the following two
conditions must be satisfied:
\begin{itemize}
\item $M^2$ should be large enough to suppress the contributions coming from
higher twists, and,

\item $M^2$ should be small enough in order to suppress the continuum and
higher state contributions.
\end{itemize}  

In the present analysis, both conditions are satisfied for the three form factors  
when $M^2$ varies in the region $1.0~GeV^2 \le M^2 \le 2.5~GeV^2$.

In Figs. (4), (5) and (6) we present the dependence of the charge $G_C$,
magnetic $G_M$ and quadrapole $G_Q$ form factors on $Q^2$, at three
different values $s_0=1.8~GeV^2$, $s_0=2.0~GeV^2$ and $s_0=2.2~GeV^2$ of the
threshold, and at the fixed value $M^2=1.0~GeV^2$ of the Borel parameter.

We observe from these figures that the dependence of the form factors on
$s_0$ is rather weak and when $s_0$ vary from $s_0=1.8~GeV^2$ to
$s_0=2.2~GeV^2$. Unfortunately, the sum rules fail working at small $Q^2$,
and hence do not allow determination of the magnetic moment of the $\rho$
meson with better accuracy compared to the results in the literature, such
as, the results predicted by the sum rules \cite{R6406,R6407},
Dyson--Schwinger based models \cite{R6415,R6416}, Covariant light--front
approach with constituent quark model \cite{R6417}, light--front formalism 
\cite{R6418} and light--front quark model \cite{R6419}. Our analysis
predicts that, starting from $Q^2=0.5~GeV^2$, the ratio $G_M(Q^2)/G_C(Q^2)$
is around $2.3$ at all values of $Q^2$. If we assume that this
behavior holds at smaller values of $Q^2$, we can conclude that 
$\mu \simeq 2.3$, which is quite close to the
prediction on the magnetic moment of the $\rho$ meson \cite{R6407}.
It should be noted here that the values of the form factors at different
values of $Q^2$ in this work are quite different compared to the predictions
of the other approaches. For example, our results for $G_C(Q^2)$ and
$G_M(Q^2)$ at all values of $Q^2$ are, approximately, two times smaller,
while $G_Q(Q^2)$ is two times larger (in magnitude) compared to the
prediction of \cite{R6419} (we use the same parametrization for the form
factors as in \cite{R6419}). Therefore, more data on $\rho$ meson is needed
in order to choose the "right" model in calculating the form factors. 

In conclusion, we have presented the results for the $\rho$ meson form
factors in the frame work of QLCSR. We have obtained that, in the region of
applicability of the method, at all values of $Q^2$, the ratio
$G_M(Q^2)/G_C(Q^2)$ is larger than $2$, more precisely, this ratio varies
around $2.3$. This result can be checked when more precise data on $\rho$
meson form factors is available.

\newpage

\section*{Figure captions}
{\bf Fig. (1)} The dependence of the charge form factor $G_C$ on $M^2$, at
six fixed values of $Q^2$, and at $s_0=2.2~GeV^2$.\\ \\
{\bf Fig. (2)} The same as in Fig. (1), but for the magnetic form factor
$G_M$. \\ \\
{\bf Fig. (3)} The same as in Fig. (1), but for the quadrapole form factor
$G_Q$. \\ \\
{\bf Fig. (4)} The dependence of the charge form factor $G_C$ on $Q^2$, at  
Three fixed values of $s_0$, and at $M^2=1.0~GeV^2$.\\ \\
{\bf Fig. (5)} The same as in Fig. (4), but for the magnetic form factor
$G_M$. \\ \\
{\bf Fig. (6)} The same as in Fig. (4), but for the quadrapole form factor
$G_Q$.

\newpage

\begin{figure}
\vskip 1.5 cm
    \includegraphics{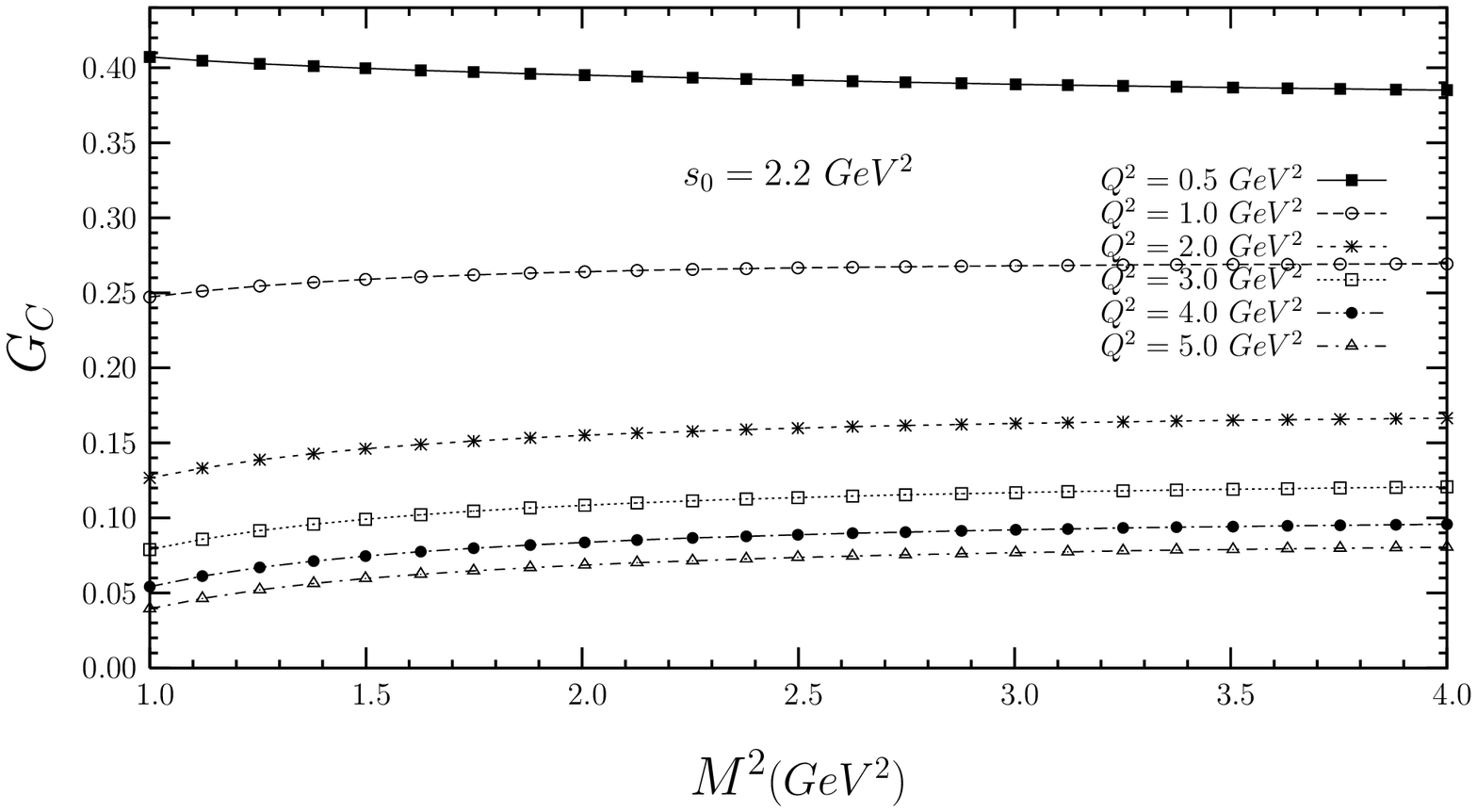}
\vskip 7.8cm
\caption{}
\end{figure}

\begin{figure}
\vskip 2.5 cm
    \includegraphics{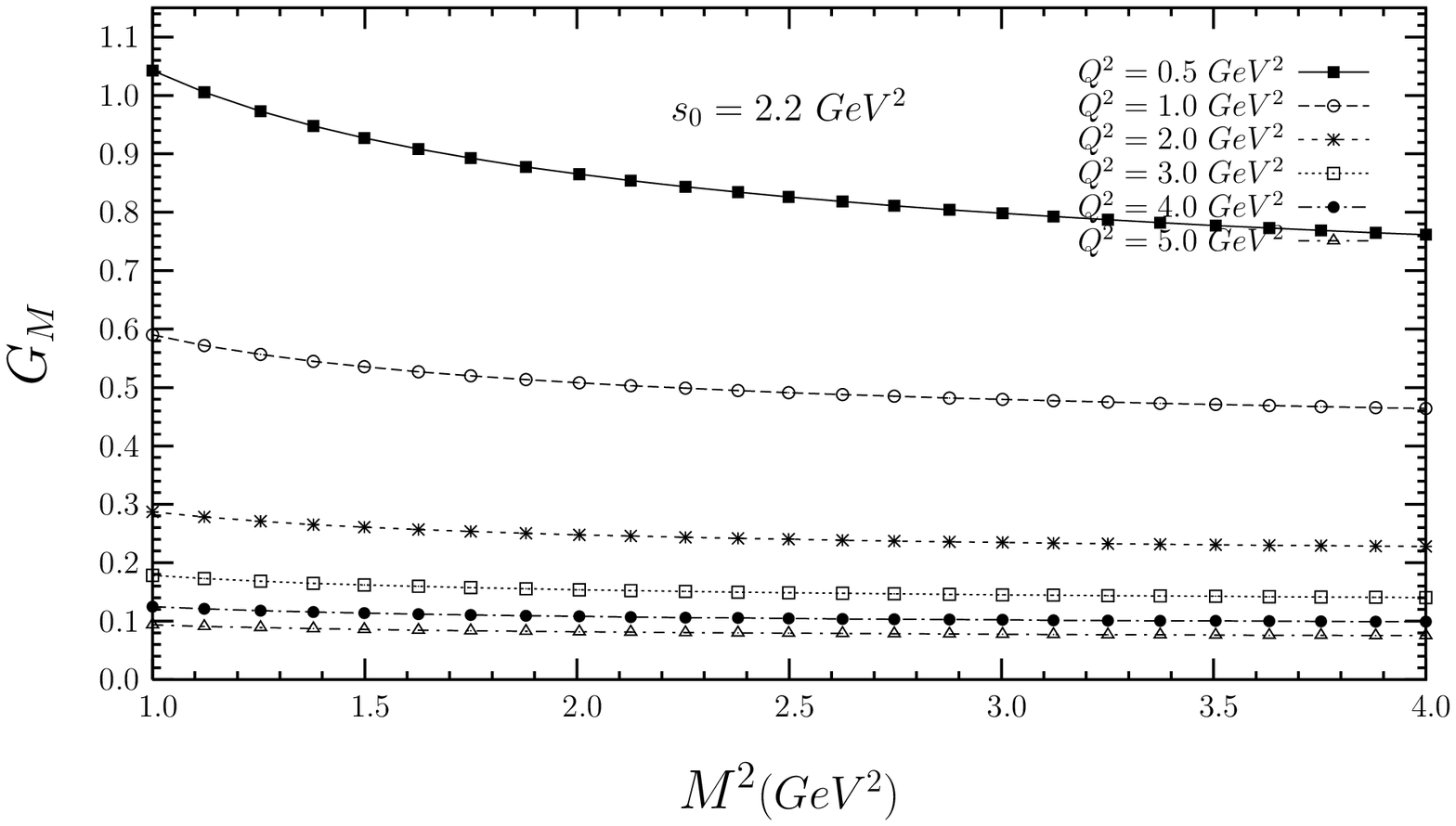}
\vskip 7.8 cm
\caption{}
\end{figure}

\begin{figure}
\vskip 1.5 cm
    \includegraphics{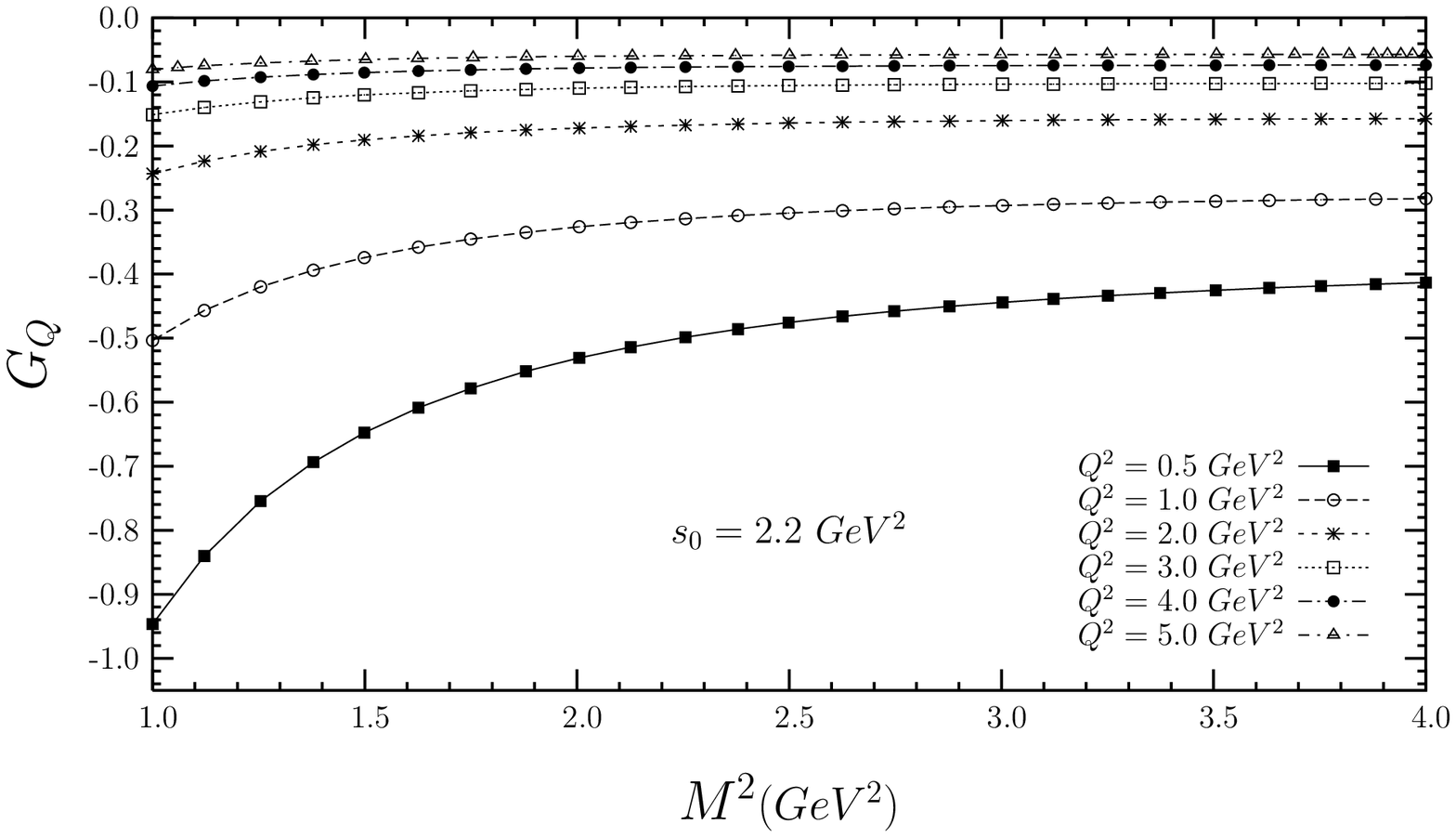}
\vskip 7.8cm
\caption{}
\end{figure}

\begin{figure}
\vskip 2.5 cm
    \includegraphics{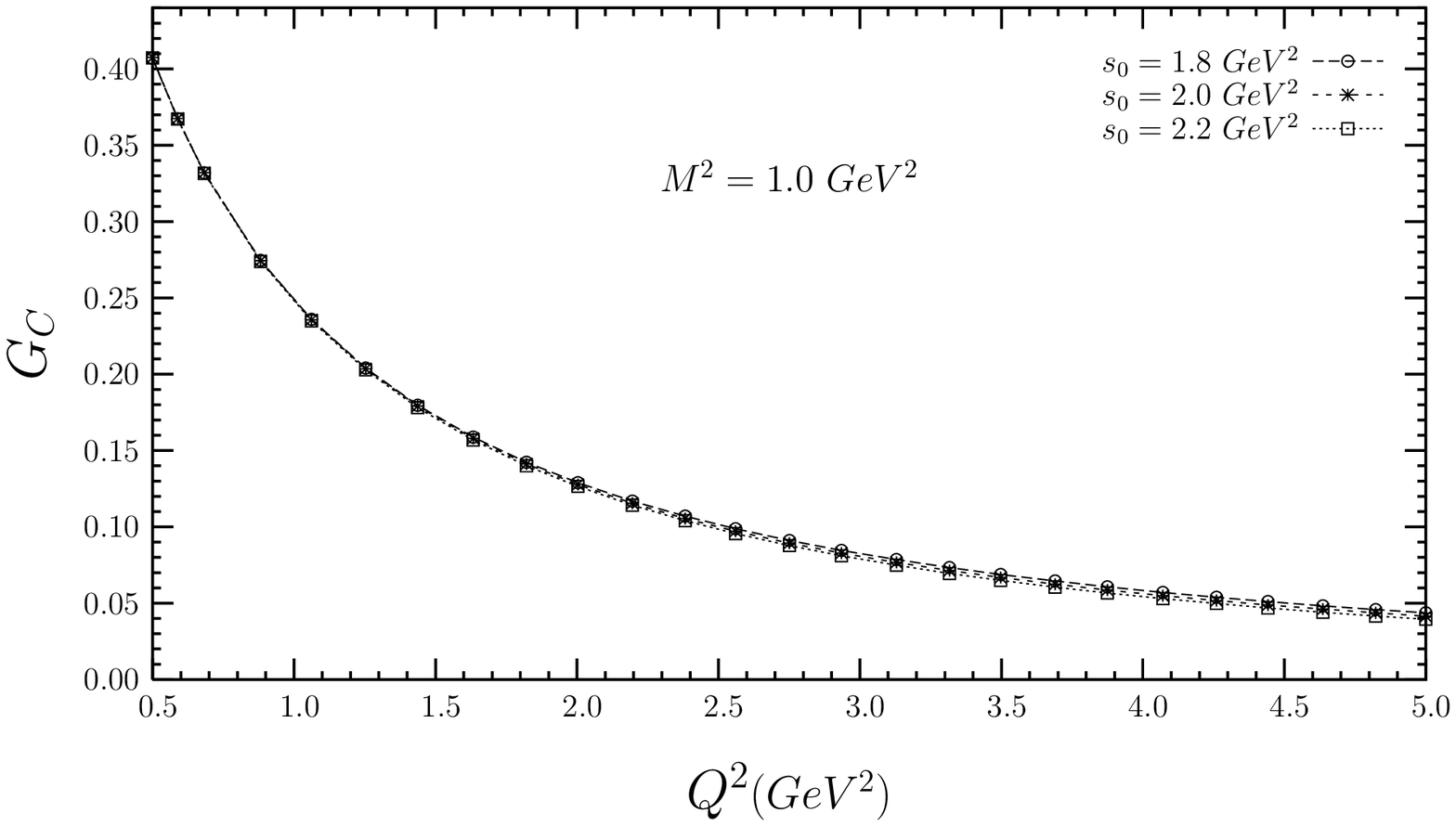}
\vskip 7.8 cm
\caption{}
\end{figure}

\begin{figure}
\vskip 2.5 cm
    \includegraphics{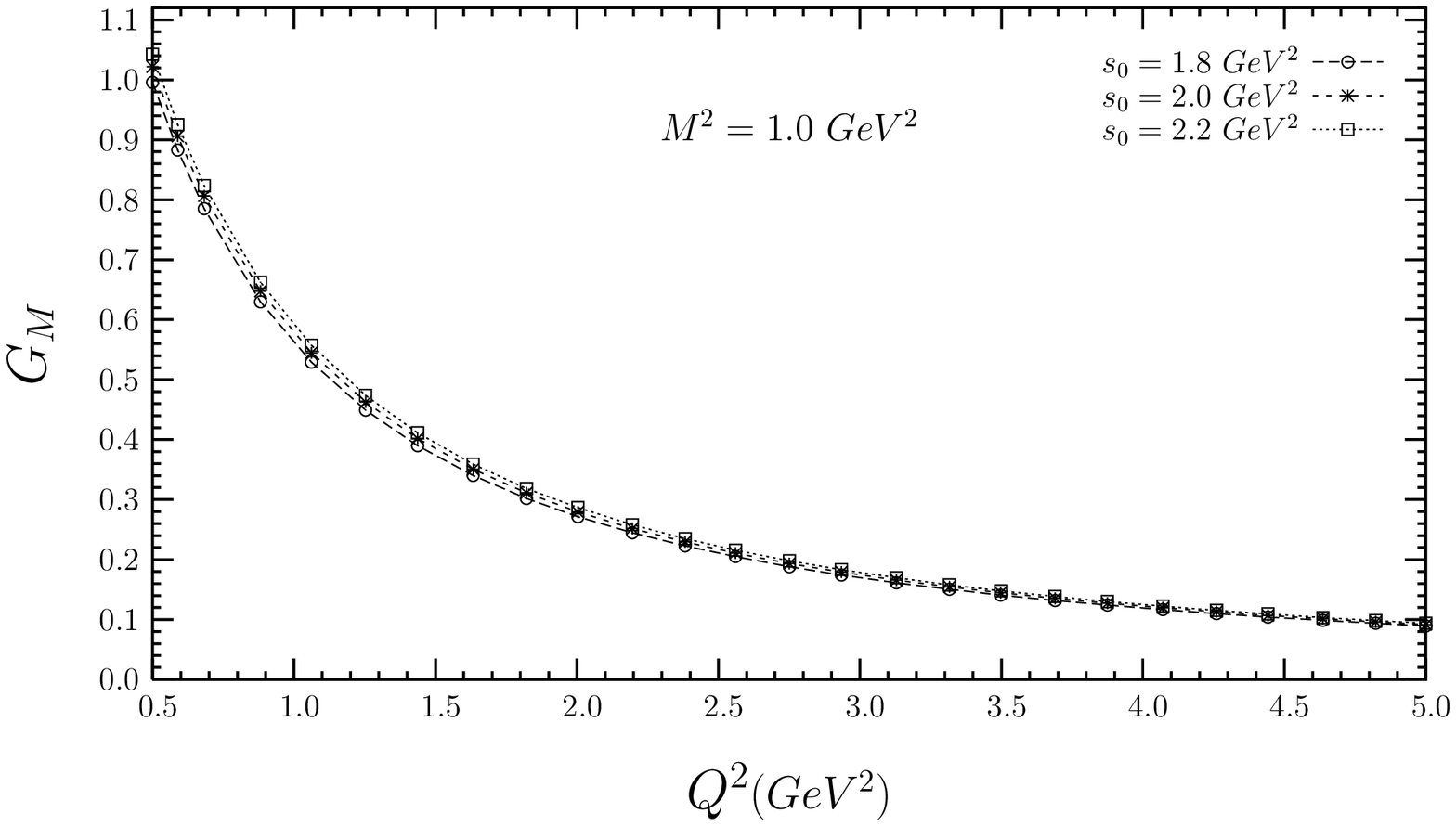}
\vskip 7.8 cm
\caption{}
\end{figure}

\begin{figure}
\vskip 1.5 cm
    \includegraphics{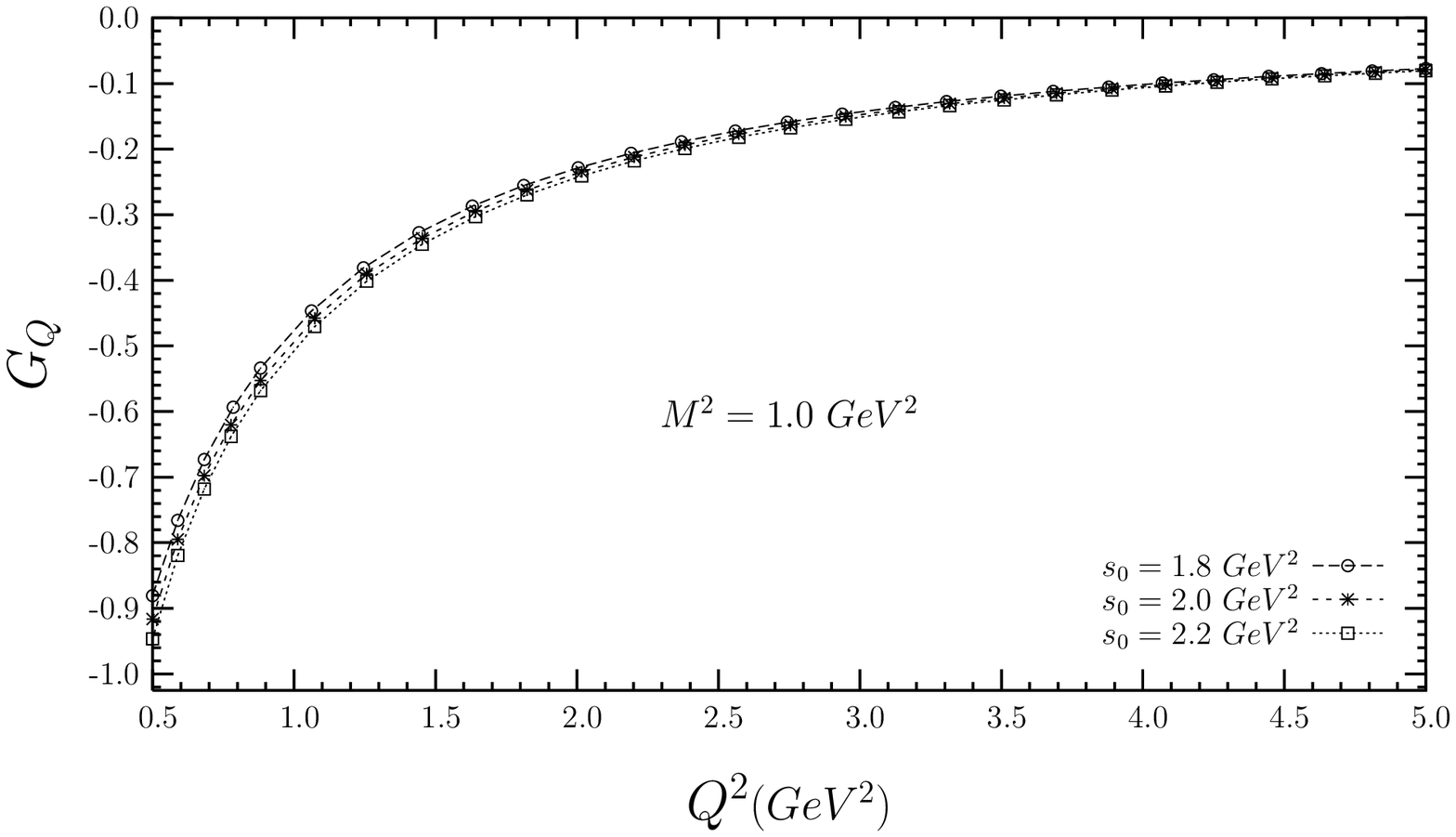}
\vskip 7.8cm
\caption{}
\end{figure}

\end{document}